\documentclass[aps,prl,preprintnumbers,amsmath,amssymb,latexsym,nofootinbib,array,enumerate,letter,twocolumn,superscriptaddress]{revtex4-1}

\usepackage{subfigure}
\usepackage{cancel}
\usepackage{here}
\usepackage{comment,braket}
\usepackage{epsf}
\usepackage{amsmath}
\usepackage{graphics}
\usepackage{amsfonts}
\usepackage{amssymb}
\usepackage{latexsym}
\usepackage{color}
\usepackage{ulem}
\input{colordvi.tex}
\usepackage{feynmp}

\usepackage[T1]{fontenc} 
\usepackage[english]{babel}
\usepackage[colorlinks=true,linkcolor=blue,urlcolor=blue,citecolor=blue]{hyperref}
\usepackage{amsmath, amssymb, amsfonts}
\usepackage{amssymb,xcolor}
\usepackage{graphicx}
\usepackage{exscale}
\usepackage{graphicx}
\usepackage{amsmath}
\usepackage{latexsym}
\usepackage{mathrsfs}
\usepackage{amsfonts}
\usepackage{amssymb}
\usepackage{float}
\usepackage{braket}
\usepackage{subfigure}
\usepackage{verbatim}
\usepackage{slashed}

\definecolor{Green}{RGB}{199,238,206}

\newcommand{\di}{\mathrm{d}}

\newcommand{\pd}{\partial} 

\newcommand{\brnc}{\overline{N^{\rm c}_{\rm R}}}

\newcommand{\td}{\text{d}}
\newcommand{\hc}{\text{ H.c. }}
\newcommand{\beq}{\begin{eqnarray}} 
	\newcommand{\eeq}{\end{eqnarray}}

\def\Hinf{H_{\rm{inf}}^{}}
\def\NR{N_{\!R}^{}}
\def\NRb{\overline{N}_{\!R}^{}}
\def\ii{{\rm{i}}}
\def\dis{\displaystyle}
\def\leqq{\leqslant}
\def\geqq{\geqslant}
\def\HH{\mathbb{H}}
\def\({\left(}
\def\){\right)}
\def\[{\left[}
\def\]{\right]}

\newcommand{\bel}[1] {\begin{equation}\label{#1}}
	\newcommand{\beal}[1] {\begin{eqnarray}\label{#1}}
		\newcommand{\be}{\begin{equation}}
			\newcommand{\ee}{\end{equation}}
		\newcommand{\bea}{\begin{array}} 
			\newcommand{\eea}{\end{array}}

		\def\vs{\vspace*{1mm}}
		
		\def\hp{\hspace*{0.3mm}}
		
		\def\hsm{\hspace*{-0.3mm}}

\begin{document}

\title{Cosmological Signatures of Neutrino Seesaw Mechanism}

\author{{Chengcheng Han}}	
\email{hanchch@mail.sysu.edu.cn}
\affiliation{School of Physics, Sun Yat-Sen University, Guangzhou 510275, China;\\
             Asia Pacific Center for Theoretical Physics, Pohang 37673, Korea}

\author{\,{Hong-Jian He}}
\email{hjhe@sjtu.edu.cn}
\affiliation{Tsung-Dao Lee Institute $\rm{\&}$ School of Physics and Astronomy, \\
Key Laboratory for Particle Astrophysics and Cosmology,\\
Shanghai Key Laboratory for Particle Physics and Cosmology,\\
Shanghai Jiao Tong University, Shanghai, China}
\affiliation{Department of Physics, Tsinghua University, Beijing, China;\\
Center for High Energy Physics, Peking University, Beijing, China}

\author{\,{Linghao Song}}
\email{lh.song@sjtu.edu.cn}
\affiliation{Tsung-Dao Lee Institute $\rm{\&}$ School of Physics and Astronomy, \\
Key Laboratory for Particle Astrophysics and Cosmology,\\
Shanghai Key Laboratory for Particle Physics and Cosmology,\\
Shanghai Jiao Tong University, Shanghai, China}

\author{\,{Jingtao You}}
\email{119760616yjt@sjtu.edu.cn}
\affiliation{Tsung-Dao Lee Institute $\rm{\&}$ School of Physics and Astronomy, \\
Key Laboratory for Particle Astrophysics and Cosmology,\\
Shanghai Key Laboratory for Particle Physics and Cosmology,\\
Shanghai Jiao Tong University, Shanghai, China}

\begin{abstract}
\noindent 
The tiny neutrino masses are most naturally explained by seesaw mechanism through singlet right-handed neutrinos, which can further explain the matter-antimatter asymmetry in the Universe.\ In this Letter, we propose a new approach to study cosmological signatures of neutrino seesaw through the interaction between inflaton and right-handed neutrinos that respects the shift symmetry.\ 
In our framework, after inflation the inflaton predominantly decays into right-handed neutrinos and its decay rate is modulated by the fluctuations of Higgs field that act as the source of curvature perturbations.\ This gives a new realization of
Higgs modulated reheating, and it produces primordial non-Gaussian signatures that can be measured by the forthcoming large-scale structure surveys.\ We find that these surveys have the potential to probe a large portion of the neutrino seesaw parameter space, opening up a new window for testing the high scale seesaw mechanism.
\\[2mm]
Phys.\ Rev.\ D\,112 (2025) L081309 (Letter) 
{[arXiv:2412.21045].}
\\[0.8mm]
(Its companion long paper: Phys.\ Rev.\ D\,112 (2025) 083555 [arXiv:2412.16033].)
\end{abstract}
			
\maketitle

\noindent \textbf{1.\,Introduction}
\vspace*{0.5mm}

Understanding the origin of tiny neutrino masses of 
$O$(0.1{\hp}eV) poses a major challenge to the standard model (SM) of particle physics.\ 
Neutrino seesaw provides the most natural explanation of tiny neutrino masses 
by including the singlet right-handed neutrinos\,\cite{Minkowski:1977sc}\cite{seesaw2},
and it further explains the matter-antimatter asymmetry (baryon asymmetry) in the Universe 
via leptogenesis\,\cite{Fukugita:1986hr}.\ 
But, the natural scale of the seesaw mechanism is around $10^{14}\hp$GeV\,\cite{foot-seesaw}
for the Higgs-neutrino Yukawa couplings of $O(1)\hp$.\ 
Probing such high scale of neutrino seesaw is truly important but extremely difficult and is far beyond the reach of current particle experiments.

\vspace*{-2mm}

In contrast, inflation provides the most appealing mechanism for dynamics of the  
early Universe, during which the Universe underwent a short period of rapid exponential expansion 
that resolves the flatness and horizon problems as well as simultaneously generating the 
primordial fluctuations for seeding the large-scale structures of the Universe.\ 
The energy scale of inflation could be as high as $10^{16}\hp$GeV, characterized by 
a nearly constant Hubble parameter $H_\text{inf}^{}$ around $10^{14}\hp$GeV, 
that coincides with the scale of neutrino seesaw.\ 
Inflation is typically driven by a scalar field known as inflaton.\ 
The primordial fluctuations arise from the inflaton's quantum fluctuations and can be 
directly measured through their contributions to the Cosmic Microwave Background (CMB).\ 
The current CMB data indicate that these fluctuations are predominantly adiabatic and Gaussian.\ 
However, during inflation the primordial perturbations could also exhibit non-Gaussianity (NG) \cite{Maldacena0210603}\cite{Meerburg:2019qqi}.\ 
The NG is sensitive to new physics effects at high energy scales.\  
Although the current CMB observations only set a weak limit on the NG parameter 
$f_{\rm NL}^{}\!\!=\!O(10)$ \cite{Planck9}, the upcoming experiments will improve 
detection sensitivity to the level of $f_{\rm NL} \!\!=\!{O}(0.01)$ 
\cite{Meerburg:2019qqi}\cite{Munoz:2015eqa}\cite{Meerburg:2016zdz}, opening up an important
window for probing the high-scale new physics.

We note that the neutrino seesaw scale $M$ 
with natural Yukawa couplings 
[$\hp y_{\nu}^{}\!\!=\hsm\!O(1)\hp$] 
is around $10^{14}\,$GeV, 
which coincides with the upper range of the inflation scale.\ 
Thus, neutrino seesaw mechanism could leave distinctive imprints in the cosmological evolution.\ It is natural to expect that the inflaton couples directly to the right-handed neutrinos and  
decays predominantly into them after inflation.\ Then the right-handed neutrinos further decay into the SM particles via Yukawa interactions, completing the reheating process.\  Moreover, during inflation the Higgs field acquires a value near the Hubble scale, varying across different horizon patches.\ This variation leads to space-dependent right-handed neutrino masses via the seesaw mechanism, which modulate the rate of inflaton decays into right-handed neutrinos.\ With these, we propose a {\it new realization of Higgs modulated reheating}, which provides a source of primordial curvature perturbations\,\cite{Dvali0303591}.\ In this Letter, 
we construct the inflaton coupling to right-handed neutrinos through an effective dimension-5 operator respecting shift symmetry.\ 
We investigate the effects of Higgs-modulated reheating and the associated NG signatures, with which we demonstrate the potential to probe the high-scale neutrino seesaw within our framework.\ 
We map the measurement of non-Gaussianity $f_\text{NL}^\text{local}$ 
onto the $(y_\nu^{},\,M)$ plane, 
which shows sensitivity to probing the light neutrino mass  
and interplays with the low energy neutrino oscillation experiments.\ 
We will further show the sensitivity of the NG measurement to 
the SM Higgs self-coupling at the inflation scale, which is quantitatively connected to the Higgs self-coupling at the TeV scale (through the renormalization group evolution).\ 
Hence, we establish the interplay between the Higgs self-coupling constraints at the inflation scale and the Higgs self-coupling measurements at the TeV scale of the LHC.

\vspace*{2.5mm}

\vspace*{1mm}			
\noindent \textbf{2.\,Dynamics of Higgs Field During and After
\\ 
\hspace*{2.6mm} Inflation} 
\vspace*{0.5mm}

During inflation, the Universe is 
effectively de Sitter spacetime.\ The dynamics of a spectator Higgs field in this de\,Sitter spacetime can be described through the stochastic inflation approach\,\cite{STAROBINSKY1982175}\cite{Starobinsky1994}.\ 
In the unitary gauge, the Higgs field is given by
$\HH\!=\hsm\!\frac{1}{\sqrt{2\,}\,}(0,\,h)^T$.\  
The potential of the SM Higgs field during inflation is $V(h)\!=\!\frac{1}{4}\lambda\hp h^4$, where its mass term can be neglected and the Higgs self-coupling $\lambda$ could have a value of $O(0.01)$ at the inflation scale within the $3\sigma$ range of the current top mass measurement\,\cite{PDG}\cite{foot0}.\
During inflation, the long-wave mode of the Higgs field value 
$h$ can be described as a classical motion with a stochastic noise:
\begin{align}
\dot{h}(\mathbf{x},t) 
= -\frac{1}{\,3H_{\rm{inf}}^{}\,}\frac{\,\partial V\,}{\partial h}+f(\mathbf{x},t)\hp,
\end{align}
where $H_{\rm{inf}}^{}$ is the Hubble parameter during inflation,
and $f(\mathbf{x},t)$ is a stochastic background and has the two-point correlation function,
\begin{equation}
\hspace*{-2.5mm}         
\langle f({\bf x}_1^{}, t_1) f({\bf x}_2^{}, t_2^{})\rangle
\!\!=\!\!
\frac{\hp H_\text{inf}^3\hp}{4\pi^2} \hp 
j_0^{}\hsm\big(\epsilon\hp a(t_1^{}) H_{\rm{inf}}^{}|
{\bf x}_{12}^{}|\big)\delta(t_1^{}\!-\hsm t_2^{}),
\end{equation}
where $j_0^{}(z) \!=\! (\sin\hsm z)/z\,$ and
${\bf x}_{12}^{}\!=\!{\bf x}_1^{}-{\bf x}_2^{}$.\  
If inflation lasts long enough, the distribution of Higgs field would eventually reach 
an equilibrium with a probability function:
\begin{equation}
\rho_{\rm{eq}}^{}(h)=
\frac{2\lambda^{1/4}}{\,\Gamma(1 / 4)\,}
\!\(\!\hsm\frac{\,2\hp\pi^2\,}{3}\!\)^{\!\!\!1/ 4}\!\!
                   \exp\!\(\!\!{\frac{\,-2 \pi^{2} \lambda h^{4}\,}{3H_\text{inf}^4}}\hsm\!\) \!. 
\end{equation}
The root-mean-square value of the Higgs field $\bar h\!=\!\sqrt{\langle h^2 \rangle\,}$ 
is derived as follows:
\begin{eqnarray}
\label{hsquaremean} 
\bar h =  
\left[\int_{-\infty}^{+\infty}\!\!\!\td h\hp h^2 \rho_{\text{eq}}^{}(h)  \right]^{\hsm\!1/2}
\!\!\simeq\hp 0.363\!
\(\!\!\frac{\,H_\text{inf}^{}~}{\,\lambda^{1/4}\,}\!\!\)\!.
\end{eqnarray}
After inflation, we consider the inflaton potential as quadratic near its minimum, the inflaton oscillates and behaves like a matter component ($w\!=\!0$).\ Consequently, the Universe expands as 
$a\!\thicksim\! t^{2/3}$, with the Hubble parameter given by $H\!=\!{2}/{(3\hp t)}$.\ 
The evolution of the super-horizon mode of the Higgs field $h$ after inflation 
is governed by the Klein-Gordon equation:
\begin{equation}
\label{evolutionofHiggs}
\ddot{h}(t)+\frac{2}{\,t\,} \dot{h}(t)+\lambda h^{3}(t)=0 \hp.
\end{equation}
Thus, for $t \hsm\!\gg\!\hsm (\sqrt{\lambda}h_\text{inf}^{})^{-1}$ and 
$h_{\rm inf}^{} \hsm\!>\!\hsm 0$, we derive the evolution of $h(t)$ as follows\,\cite{Long}:
\begin{equation}
\label{h0 theoretical solution formal}
h(t)=
A\!\(\!\frac{\,h_\text{inf}^{}\,}{\,\lambda\,}\!\)^{\!\hsm\frac{1}{3}}\!\!t^{-\frac{2}{3}}\!\cos\hsm\!\Big(\!\lambda^{\frac{1}{6}}h_\text{inf}^{\frac{1}{3}}\hp\omega \hp t^{\frac{1}{3}}\!+\theta\hsm \Big),~~
\end{equation}
where $h_\text{inf}$ is the Higgs field value at the end of inflation which varies in different Hubble patches, 
and the parameters $(A,\,\omega,\,\theta)$ are given by
\begin{align} 
\hspace*{-2mm}
A & = \!\(\!\frac{2}{\,9\,}\!\)^{\!\!\frac{1}{3}}\! 5^{\frac 1 4} 
\!\simeq 0.9\hp,~~~
\dis\omega =\frac{\,\Gamma^2(3/4)\,}{\sqrt{\pi\,}}
{6}^{\frac{1}{3}} 5^{\frac 1 4}
\simeq 2.3\hp, 
\nonumber\\
\hspace*{-2mm}
\theta & = -3^{-\frac 1 3}2^{\frac 1 6}\dis\omega\!-\hsm\arctan 2\simeq\! -2.9 \hp.
\end{align}
The solution can be readily generalized to the case of $h_\text{inf}^{}\!<\!0\,$.\ 
Eq.\eqref{h0 theoretical solution formal} shows that 
after inflation, the Higgs field oscillates in its quartic potential  
$\frac{1}{4}\lambda\hp h^4$, but its oscillation amplitude will gradually decrease over time~\cite{Lu:2019tjj}. 
			
\vspace*{3mm}
\noindent \textbf{3.\,Inflaton-Neutrino Interaction and Inflaton \\ 
\hspace*{2.6mm} Decay}
\vspace*{1.5mm}

The right-handed neutrinos $N_R^{}$ can couple to the inflaton $\phi$ through a unique dimension-5 effective operator which has a cutoff scale $\Lambda$ and respects the inflaton's shift symmetry\,\cite{foot-1}.\ 
Thus, we construct the minimal model incorporating 
both inflation and neutrino seesaw with the following Lagrangian:
{\small 
\begin{align}
\label{Total_Lagrangian}
\hspace*{-2mm}
& \Delta \mathcal{L} =\sqrt{-g\,}\hp\bigg[\!
-\hsm\frac{1}{2}\hp\partial_{\mu}\phi\hp\partial^{\mu}\phi -\!V(\phi) 
+ \hp \NRb\hp {\ii} \slashed{\pd}\hp \NR 
\\
\hspace*{-2mm}
&+\hsm\frac{1}{\Lambda} \pd_\mu \phi\, \NRb \gamma^\mu \gamma^5 \hsm\NR    
\!+\!\hsm\(\!\hsm-\frac{1}{2} M \brnc \NR\!- y_\nu  
\overline{L}_{\rm L} \widetilde{\mathbb{H}}\NR \hsm  +\!\!\hc\!\!\!\hsm\)\! \!\bigg], 
\nonumber 
\end{align} 
}
\hspace*{-2.6mm}
where $V(\phi)$ is the inflaton potential and 
$L_{\rm L}\!=\! (\nu_{\rm L}^{},\hp e_{\rm L}^{})^T$ 
is the left-handed lepton doublet.\  
After inflation we consider the inflaton mass term dominates 
the potential $V(\phi)$ under which the inflaton $\phi$ will oscillate.\ 
Due to the shift symmetry, inflaton couples to the right-handed neutrinos through the dimension-5 effective operator of Eq.\eqref{Total_Lagrangian}\,\cite{foot-UVmodel}.\ 
The perturbative unitarity imposes a lower bound on its cutoff scale,  
$\Lambda\!\gtrsim\hsm 60 H_{\rm inf}$.\ 
In our setup, this dimension-5 operator causes the inflaton to decay predominantly into right-handed neutrinos after inflation.\  If the inflaton couples to 
the SM fermions via dimension-5 operators and with the shift symmetry, 
the corresponding decay rates are suppressed by the fermion masses, 
as the Higgs field quickly decreases after inflation.\ 
(This also applies to the case of inflaton 
coupling to top quarks.)\
Couplings between the inflaton and SM gauge bosons (via operators such as $\phi F^{\mu\nu} \tilde{F}_{\mu\nu}$) can be forbidden if the shift symmetry is anomaly-free with respect to the SM gauge group, i.e., the sum of the anomaly parts of fermion triangle loops (containing $\phi$ and two SM gauge bosons as external lines) vanishes and thus ensure inflaton to mainly decay into the right-handed neutrinos.

For simplicity, we focus on analyzing the case of one family of fermions.\ 
For the neutrino seesaw with $|y_\nu^{}h| \!\ll\! M$, the two mass eigenstates 
$\nu$ and $N$ have masses:
\begin{equation}
\label{eq:Mass-nu-N}
m_{\nu}^{}\!=\! -\frac{~y_\nu^2 h^2\,}{2 M},~~~ 
M_{N}^{} \!=\! M\!+\!\frac{~y_\nu^2 h^2\,}{2M}.
\end{equation}
The rotation angle $\theta$ for this mass-diagonalization is given by  
$\tan\hsm\theta \!\simeq\! {y_\nu^{}h}\hsm/(\hsm\sqrt{2\,}M)\,$.\ 
In Eq.\eqref{eq:Mass-nu-N}, the heavy neutrino mass $M_N^{}$ has a shift
$y_\nu^2 h^2/{(2M)}$ from $M$, which is crucial for our mechanism as we are really 
probing the seesaw effect on the heavy neutrino mass eigenvalue.\ 

\vspace*{1mm}

For $|y_\nu^{}h| \!\ll\! M$, the inflaton decay rate into 
neutrinos is given by
\begin{equation}
\label{decay_rate}
\Gamma\simeq\frac{\,m_{\phi}^{}M^2\,}{4\pi\Lambda^2}\!\!
\left[\hsm 1\!+\frac{1}{4}\!\(\!\!\frac{\,y_\nu^{}h\,}{M}\!\!\)^{\!\!2}
\right]\!.
\end{equation}
where kinetic factors are ignored for simplicity, 
but $m_\phi^{} \!\!>\!\! 2 M_N$ 
is always required to ensure that the inflaton decay channel
$\phi\hsm\to\hsm NN$ is kinematically open.\ 
We see that the inflaton decay rate depends on the Higgs field value $h\hp$ \cite{foot-x}.\ 
On the other hand, Refs.\,\cite{Arkani-Hamed:2015bza}-\cite{Hook:2019zxa} studied cosmological collider (CC) signals, which manifest as oscillatory features in the primordial NG\,---\,a qualitatively different type of NG signal from what we study.\ 
In particular, Ref.\,\cite{Chen:2018xck} explored the CC signal arising from a coupling similar to our Eq.\eqref{Total_Lagrangian}, this would provide a complementary probe within our model framework.

\vspace*{3mm}

\noindent \textbf{4.\,Curvature Perturbation from Higgs Modulated  
\\
\hspace*{2.6mm} Reheating}
\vspace*{1.5mm}

In our approach, the inflaton decay rate is affected by the value of the SM Higgs field.\ 
The variation of the Higgs field $h(\mathbf{x}, t_\text{reh}^{})$ leads to a spatial variation 
of the decay rate $\Gamma_\text{reh}^{}(\mathbf{x})$.\ 
It perturbs the local expansion history, seeding large-scale inhomogeneity and anisotropy.\ 
These fluctuations can be described by the $\delta N$ formalism\,\cite{Starobinsky:1982ee, Salopek:1990jq, Comer:1994np, Sasaki:1995aw, Sasaki:1998ug, Wands:2000dp, Lyth:2003im, Rigopoulos:2003ak, Lyth:2004gb}.\ 
The number of e-folds of the cosmic expansion 
after inflation can be computed as\,\cite{foot2}{\hp}:
\begin{align}
\hspace*{-7mm}
N({\bf x}) 
&= \int\!\! \di\hsm \ln a(t)
=\int\limits_{t_{\rm end}}^{t_{\rm reh}({\bf x})}\!\!\!
\!\! 
\di t\hp H(t)\hp +\!\!\int\limits_{t_{\rm reh}({\bf x})}^{t_{\rm f}}\!\!\!\!\!
\di t\hp  H(t)   \nonumber
\\
\hspace*{-7mm}
&=\!\!\!
\int\limits_{\rho_{\rm end}}^{\rho_{\rm reh}(h({\bf x}))}\!\!\!\!\!
\hspace*{-1mm}
\di\rho\hp\frac{\,H\,}{\dot{\rho}} +\!\!\!\!\!
\int\limits_{\rho_{\rm reh}(h({\bf x}))}^{\rho_{\rm f}}\!\!\!\!
\hspace*{-2mm}
\di\rho\hp\frac{\,H\,}{\dot{\rho}} \,,
\end{align}
where $\hp a(t)$ is the scale factor and $\rho(t)$ is the total energy density of the Universe 
at the time $t$.\ 
The curvature perturbation during reheating, $\zeta(\mathbf{x},t)$, is 
equal to the $\delta N(\mathbf{x},t)$ of cosmic expansion among different Hubble patches 
in the uniform energy density gauge:
\begin{equation}
\zeta_h(\mathbf{x},t)=\delta N(\mathbf{x},t)=N(\mathbf{x},t)\!-\!\langle N(\mathbf{x},t) \rangle\hp.
\end{equation}
			
For this study, we describe the Universe as a perfect fluid both before 
and after the completion of reheating.\ 
During the period $t_\text{end} \!<\! t \!<\! t_\text{reh}$, 
we consider the inflaton potential is dominated by its mass term.\ 
Thus, when the inflaton oscillates near the minimum of the potential, the Universe is 
matter-dominated ($w\!=\!0$).\ (Our approach also applies to the general case of 
$w\!\neq\!{1}/{\,3\,}$.)\ For the period $t \hsm\!>\! t_\text{reh}^{}$, 
consider the right-handed neutrinos decay fast enough after being produced,
so the Universe transitions to a radiation-dominated phase ($w\!=\!{1}/{\,3\,}$).\ 
Using the equation of state $\dot{\rho} \hsm +\hsm 3H(1\!+\!w)\rho \!=\! 0\hp$, 
the locally expanded e-folding number can be expressed as follows:
\begin{equation}
N({\bf x}) 
=-\frac{1}{3}\ln\! \frac{\,\rho_{\rm reh}\hsm\big(h({\bf x})\hsm\big)\,}{\rho_\text{inf}}
\!-\!\frac{1}{4}\ln\! \frac{\rho_f}{\,\rho_{\rm reh}\hsm\big(h({\bf x})\hsm\big)\,}\,.
\end{equation}
Using first Friedmann equation $3H^2M_{p}^2 \!\!=\!\!\rho\,$, 
and noting that reheating completes when 
$H(t_\text{reh}^{})\!\!=\!\Gamma_\text{reh}^{}\hp$
(where we take the sudden reheating approximation), 
we determine the curvature perturbation after reheating 
($t \!>\hsm t_\text{reh}^{}$){\hp}:
\begin{align}
\zeta_h^{}({\bf x},t \!>\! t_\text{reh}^{})
&=\delta N({\bf x})=N({\bf x})\!-\!\langle N({\bf x})  \rangle \nonumber
\\
&=-\frac{1}{12}\Big[\!\ln \rho_{\rm reh}({\bf x})\!-\!
\langle\ln\rho_{\rm reh}({\bf x})\rangle \Big]  \nonumber
\\
&=-\frac{1}{6}\big[\!\ln(\Gamma_\text{reh})\!-\!\langle \ln(\Gamma_\text{reh})\rangle \big]\hp.
\end{align}

Combined with the inflaton fluctuation $\delta \phi$ during inflation, 
the total comoving curvature perturbation is given by
$\zeta\!=\!\zeta_\phi^{} \!+\hsm \zeta_h^{}\hp$, 
where $\zeta_\phi^{}$ is generated by the inflaton fluctuation  $\delta \phi\hp$,
\beq 
\zeta_\phi^{} \,\simeq\, -\frac{\,H_\text{inf}^{}\,}{\dot{\phi_0}} \delta \phi(\mathbf{x}) \,, 
\eeq 
and $\zeta_h^{}$ originates from the effect of Higgs-modulated reheating.\ 
Because these two components are generated at different times and are independent of 
each other, the power spectrum of $\,\zeta\hp$ contains both contributions:  
\beq 
\mathcal{P}_{\zeta} = \mathcal{P}_{\zeta}^{(\phi)} + \mathcal{P}_{\zeta}^{(h)},
\eeq
where $\mathcal{P}_{\zeta}^{(\phi)}$ is the contribution of inflaton fluctuations,
\beq 
\mathcal{P}_{\zeta}^{(\phi)} = 
\(\hsm\!\frac{\,H_\text{inf}\,}{\hp\dot{\phi}\,}\!\)^{\!\!2} \!\mathcal{P}_\phi^{} =\(\!\hsm\frac{\,H_\text{inf}\,}{\hp\dot{\phi}\,}\!\)^{\!\!2}\!\frac{H_\text{inf}^2}{\,4\pi^2\,}\,.
\eeq 
We further define $R$ as square root of the ratio between the power spectra 
of Higgs-modulated reheating and of the comoving curvature perturbation $\zeta\,$,
%

\begin{align}
\nonumber
\\[-13mm]
\label{rh}
R \,\equiv\(\!\!\frac{~\mathcal{P}_{\zeta}^{(h)}}{\mathcal{P}_{\zeta}^{(o)}}\!\)^{\!\!\!1/2} ,
\end{align}
\\[-3mm]
where 
$\mathcal{P}_{\zeta}^{(o)}\!\simeq\! 2.1\!\times\! 10^{-9}$ 
is the observed curvature perturbation \cite{Planck1}\cite{Planck6}.\ 
To agree with the observation, we should require $R\! < \!1\hp$.
			
Modulated reheating can also provide a source of primordial NG.\ The primordial NG from 
the three-point correlation function of $\zeta$ is known as the bispectrum 
$\left\langle \zeta_{\mathbf{k}_{1}}^{} \zeta_{\mathbf{k}_{2}}^{} 
 \zeta_{\mathbf{k}_{3}}^{} \right\rangle$.\ 
To compute the $n$-point correlation function of $\zeta_h$, 
we expand the curvature perturbation: 
\begin{equation}
\zeta_h^{}=\delta N
=N^{\prime} \delta h_\text{inf}^{}+\frac{1}{2} N^{\prime\prime}(\delta h_\text{inf}^{})^2 
+\cdots ,
\end{equation}
where $N^{\prime}$ and $N^{\prime\prime}$ denote the first and second derivatives of the e-folding number $N$ with respect to $h_\text{inf}^{}\hp$.\  
The expansion allows us to determine the amplitude of the curvature perturbations as 
$\mathcal {P}^{(h)}_\zeta \!\!=\hsm\! {N^\prime}^2 \mathcal {P}_{h_\text{inf}}^{}$ 
and the primordial local NG $f_\text{NL}^\text{local}$
\cite{Wands10040818,Ichikawa:2008ne, DeSimone12106618,Karam210302569,Litsa201111649}.
			
We note that when reheating occurs, the value of the Higgs field is an oscillatory function of its initial value,
\begin{equation}
h(t_\text{reh}^{},h_\text{inf}^{}) \propto h_\text{inf}^{\frac{1}{3}}
\cos (\omega_\text{reh}^{} h_\text{inf}^{\frac{1}{3}} +\theta) ,
\end{equation}
where the oscillating frequency is given by 
\begin{equation}
\label{omega_reh_def}
\omega_\text{reh}^{} = \lambda^{\frac{1}{6}}t_\text{reh}^{\frac{1}{3}}\omega \,.
\end{equation}
When $t_{\rm reh}^{}$ is large, the oscillation frequency can become very high.\ 
Note that $\zeta_h^{}$ can be expanded into the form 
$A\!+\!\! B h^2/M^2\!\!+\!{O}(h^4/M^4)$,  
which includes a factor $\cos^2(\omega_{\rm {reh}} h_\text{inf}^{{1}/{3}}\!+\!\theta)\hp$.\ 
Since $h_\text{inf}^{}$ varies across different Hubble volumes and $\zeta_h^{}$ 
is highly sensitive to $h_\text{inf}^{}\hp$, 
averaging over a sufficiently large volume 
makes the factor $\cos^2(\omega_{\rm {reh}} h_\text{inf}^{{1}/{3}}\!+\!\theta)$ 
be effectively as ${1}/{2}$ \cite{Suyama:2013dqa}.\ 
Thus, for the following analysis we set 
$\!\cos^2(\omega_{\rm {reh}} h_\text{inf}^{{1}/{3}}\!+\!\theta)\! 
\!\rightarrow\! {1}/{2}\,$ in the expression of $\hp\zeta_h^{}\hp$.

\vspace*{4.5mm}	
\noindent 
\textbf{5.\,Bispectrum from Higgs Fluctuations}
\vspace*{4.5mm}

We expand the curvature perturbation 
in terms of the Higgs fluctuation as follows:
\label{zetaHiggsAppendix}
\begin{align}
\zeta_h(\mathbf{x})& = z_1^{}\delta h_\text{inf}^{}(\mathbf{x})\!+\!
\frac{1}{\,2\,} z_2^{} \delta h_\text{inf}^2(\mathbf{x})\hp,
\end{align}
where the coefficients $z_1^{}$ and $z_2^{}$ are given by 
\begin{equation}
z_1^{}=-\left.\frac{\,\Gamma'\, }{\,6\Gamma\,}
\hsm\right|_{h_{\mathrm{inf}}=\bar{h}}\!,
~~~
z_2^{}=\!\left.\frac{\,\Gamma'\Gamma'\!-\hsm\Gamma \Gamma''\,}{6\hp\Gamma^2}\hsm\right|_{h_{\mathrm{inf}}=\bar{h}}\!,
\end{equation}
with $\Gamma'$ ($\Gamma''$) being the first (second) derivative of $\Gamma$ respect to 
$h_{\rm \inf}^{}$. 
In the following, we  abbreviate the Hubble scale during inflation $H_\text{inf}^{}$ 
as $H$ and the Higgs field value during inflation $h_\text{inf}^{}$ as $h$
which should differ from the Higgs field value $h(t)$ after inflation.\
The three-point correlation function of $\hp\zeta\hp$ from modulated reheating 
$\langle \zeta_{\mathbf{k}_1}^{}\zeta_{\mathbf{k}_2}^{}\zeta^{}_{\mathbf{k}_3}\rangle_h^{}$ 
consists of two parts: 
\\[-4.5mm]
\begin{align}
\hspace*{-1.5mm}
\langle \zeta_{\mathbf{k}_1}^{}\zeta_{\mathbf{k}_2}^{}\zeta_{\mathbf{k}_3}^{}\rangle_h^{}
\!=\! z_{1}^3\langle \delta h_{\mathbf{k}_1}\delta h_{\mathbf{k}_2} \delta h_{\mathbf{k}_3}
\rangle \!+\! z_1^2z_2^{}\langle\delta h^4\rangle\hsm 
(\mathbf{k}_1^{}, \mathbf{k}_2^{}, \mathbf{k}_3^{}) \hp.
\label{3ptz_h}
\end{align}
\\[-3.5mm]
On the right-hand side of the equality in Eq.\eqref{3ptz_h}, the first term 
is the three-point correlation function of the Higgs fluctuation $\delta h(\mathbf{k})$ 
generated by the self-interactions of Higgs field, whereas the second term arises from 
replacing one $\delta h(\mathbf{k})$ by the nonlinear term 
$\frac{1}{2}z_2\delta h^2$, which exists even if the Higgs fluctuation $\delta h(\mathbf{k})$ 
is purely Gaussian.

\vs 
            
During inflation, the Higgs fluctuation $\delta h$ could be treated as a nearly massless scalar.\ 
Due to the SM Higgs self-coupling term 
$\Delta\mathcal{L}\!\!\!=\!\!\!-\sqrt{-g\,}[(\lambda {\hp}\bar{h})\delta h^3]$ 
and according to the Schwinger-Keldysh (SK) path integral formalism\,\cite{Weinberg2005}\cite{Chen170310166}, 
the three-point correlation function of the Higgs fluctuation $\delta h$, 
$\langle \delta h_{\mathbf{k_1}}\delta h_{\mathbf{k_2}}\delta h_{\mathbf{k_3}}\rangle'$ \cite{foot3},  
can be computed through the following integral:
%
\begin{equation}
\label{3pth_SM0}
\hspace*{-2.5mm}
\langle \delta h_{\mathbf{k_1}}^{} \delta h_{\mathbf{k_2}}\delta h_{\mathbf{k_3}}\rangle'
\!=\! 12\lambda \bar{h} {\mathrm{Im}\!\hsm\(\!\int_{-\infty}^{\tau_f} \hspace*{-2.5mm}
\di\tau\hp a^4 \!\prod _{i=1}^3\! G_{+}(\mathbf{k}_i^{},\tau)\!\!\)}\hsm,
\end{equation}
where $G_{\pm}(\mathbf{k}_i,\tau)$ is the bulk-to-boundary propagator of massless scalar 
in the SK path integral\,\cite{Chen170310166}.\ 
In Eq.\eqref{3pth_SM0}, we denote the integral part $\rm{Im}(\cdots)\!\equiv\!A$
and compute it to the leading order of $k_t^{}\tau_f^{}\hp$:  
\begin{align}
\label{int-A1-A}
\hspace*{-4mm}
A =& \frac{H^2}{\,24k_1^3k_2^3k_3^3\,}
\!\biggl\{\!(k_1^3\!+\!k_2^3\!+\!k_3^3)\!\Big[\!\ln(k_t|\tau_f|)\!+\!\gamma\!-\!\frac{4}{3}\Big]
\nonumber
\\
\hspace*{-4mm}
& +k_1k_2k_3\!-\!\sum_{a\neq b}\!k_a^2k_b\!
\biggr\},
\end{align}
where $\hp\gamma\hsm\simeq\hsm 0.577\hp$ is the Euler-Mascheroni constant, 
and the wavenumber $\,k_t^{}\hsm\!=\! k_1^{}\!+\hsm k_2^{}\!+\hsm k_3^{}$
is around the scale of the present observable Universe.\  
Here $H\equiv H_{\rm inf}$ for simplicity. 
For the second term on the right-hand side of Eq.\eqref{3ptz_h}, 
it can be expressed as a 4-point correlation function of $\delta h$, 
and to the leading order it is given by the product of two 2-point correlation functions:
\begin{equation}
\begin{aligned}
& z_1^2z_2^{}\langle\delta h^4\rangle(\mathbf{k_1},\mathbf{k_2},\mathbf{k_3})
\\
& =(2\pi)^3\delta^3(\mathbf{k_1}\!+\!\mathbf{k_2}\!+\!\mathbf{k_3})
z_1^2z_2^{}\!\hsm\(\!\!\frac{H^4}{\,4k_1^3k_2^3\,}\!+\! 2\,\text{perm.}\!\)\!.~~
\end{aligned}
\end{equation}

For this study, we mainly focus on the magnitude of local NG predicted by our model, 
which can be approximated as follows:
\begin{equation}
\label{fnllocal_appro}
f_{\mathrm{NL}}^{\mathrm{local}} \simeq 
-\frac{10}{3}\frac{z_1^3H^3}{\,(2\pi)^4 \mathcal{P}_{\zeta}^{2}\,}\!
\(\!\!\frac{\,\lambda\hp\bar{h}\,}{\,2H\,}N_e \!-\!\frac{\,z_2^{}H \hp\,}{4z_1^{}}\!\)\!.
\end{equation}
where $N_e$ is the e-folding number corresponding to the present Universe,
\begin{equation}
N_e \!= \ln\!\frac{\,a_\text{end}^{}\,}{a_k^{}}
=\ln\!\frac{\,-(\hsm H\tau_f^{}\hsm)^{-1}\,}{\,{k_t^{}}/{H}\,}
\!=\!-\ln (k_t^{}|\tau_f|) \!\thicksim 60 \hp.~~ 
\end{equation}
We find that the contribution from the Higgs self-coupling can be dominant (not studied before), 
whereas nonlinear term contribution is non-negligible.

\vspace*{6.5mm}			
\noindent 
\textbf{\hspace*{-2.4mm}6.\,Probing\,Neutrino\,Seesaw Using\,Non-Gaussianity}
\vspace*{1mm}

\begin{figure}[t]
\vspace*{-2mm}
\centering
\includegraphics[width=0.48\textwidth]{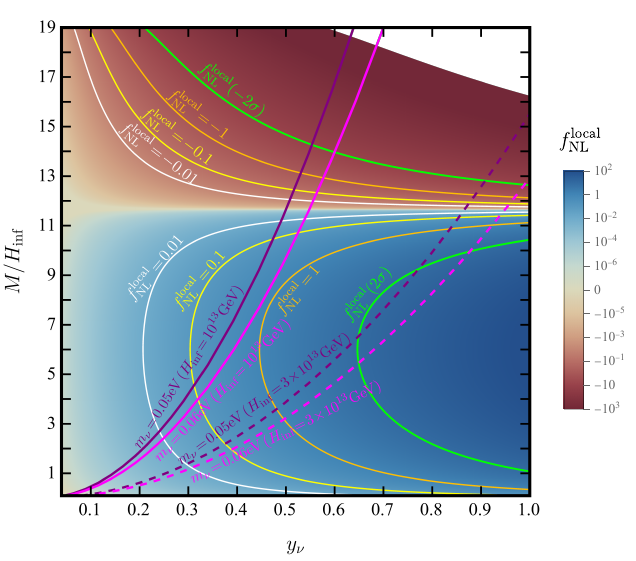}
\vspace*{-8mm}
\caption{Prediction of the non-Gaussianity (NG) $f_{\rm{NL}}^{\rm{local}}$ 
from the seesaw parameter space of the heavy neutrino mass scale $M$ 
versus Yukawa coupling $y_\nu^{}\hp$.} 
\label{fig:1}
\vspace*{-2mm}
\end{figure}

\begin{figure}[t]
\vspace*{-2mm}
\centering
\includegraphics[width=0.48\textwidth]{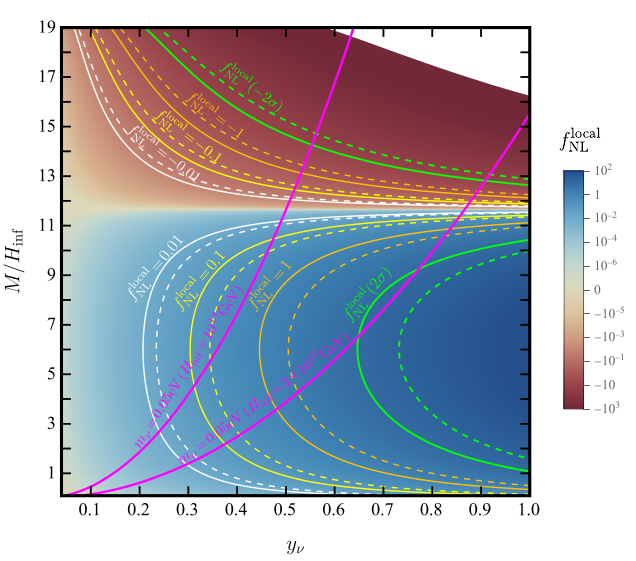}
\vspace*{-8mm}
\caption{Prediction of the non-Gaussianity $f_{\rm{NL}}^{\rm{local}}$ 
from the seesaw parameter space of heavy neutrino mass scale $M$ 
versus the Yukawa coupling $y_\nu^{}\hp$, 
where we input the SM Higgs self-coupling  
$\lambda\!=\!0.01\hp$ (solid curves) and $\lambda\!=\!0.02\hp$ (dashed curves), 
and the Hubble parameter during inflation is set as 
$H_\text{inf}^{}\!=\!10^{13}\text{GeV}$ and 
$3\!\times\! 10^{13}\text{GeV}$ respectively.} 
\label{fig:2}
\vspace*{-2mm}
\end{figure}

In our analysis, 
the amplitude of the comoving curvature perturbation power spectrum 
$\mathcal{P}_{\zeta}$ is taken as, 
$\ln(10^{10} \mathcal{P}_{\zeta}) \!\!\simeq\!\! 3.047$, 
according to the Planck-2018 data \cite{Planck1}\cite{Planck6}.\  
We set the SM Higgs self-coupling $\lambda=0.01$, and 
the Hubble parameter $H_{\rm{inf}}^{}\!=\!10^{13}\hp$GeV 
(or, $3\times\! 10^{13}\hp$GeV).\ 
We also set the inflaton mass $m_\phi^{}\!\!=\!40 H_{\rm inf}^{}$ and 
the cutoff scale $\Lambda\!=\!60 H_{\rm inf}^{}\hp$.\ 
(The effects due to the variation of inputs will be shown 
in Table\,\ref{tab:1}.)
With these inputs, we present our findings 
in Figs.\,\ref{fig:1} and \ref{fig:2}.

In Fig.\,\ref{fig:1}, the colored region obeys the condition $R \!<\! 1\hp$
and the white region in the upper-right corner corresponds to $R\!\geqq\! 1\hp$.\ 
The region with blue color corresponds to  
$f_{\mathrm{NL}}^{\mathrm{local}}\!\!>\hsm\!0\hp$, 
whereas the red regions represent  
$f_{\mathrm{NL}}^{\mathrm{local}}\!\!<\!0\hp$.\ 
The green contours describe the 2$\hp\sigma$ bounds on 
$f_{\rm{NL}}^{\rm{local}}$ from Planck-2018 data, 
$-11.1 \!\!\leqq\! f_{\mathrm{NL}}^{\mathrm{local}}\!\leqq\! 9.3\hp$ \cite{Planck9}.\ 
We further present contours for $f_{\rm{NL}}^{\rm{local}}\!=\!\pm 1,\,\pm 0.1,\,\pm 0.01$, 
which are plotted as orange, yellow, and white curves, respectively.\  
These contours represent sensitivity reaches by the future experiments.\  
We see that the local-type NG measurements for $f_{\rm{NL}}^{\rm{local}}\!>\!0\hp$
and  $f_{\rm{NL}}^{\rm{local}}\!<\!0\hp$ can probe different seesaw parameter space 
of the $(y_\nu^{},M)$ plane, so their probes are complementary.\  

\vs 

In Fig.\,\ref{fig:1}, we set two benchmarks for the light neutrino mass 
$m_\nu^{}\!\!=\!0.06{\hp}$eV and 0.05{\hp}eV\,\cite{foot-nu}, 
shown as the pink and purple curves respectively, 
for the Hubble parameter $H_{\rm{inf}}^{}\!=\!10^{13}\hp$GeV (solid curves) and 
$3\!\times\! 10^{13}\hp$GeV (dashed curves).\ 
We see that a larger Hubble parameter shifts the pink and purple curves 
towards the regions with larger Yukawa coupling $y_\nu^{}\hp$.\ 
For the case with 
$f_{\mathrm{NL}}^{\mathrm{local}}\!\!>\!\!0\hp$, 
we see that for a light neutrino mass 
$m_\nu^{}\!=\! 0.06\,(0.05)\hp$eV 
and $H_{\rm{inf}}\hsm\!=\hsm\! 3\!\times\!10^{13}\hp$GeV,
the existing Planck-2018 data already excluded part of the parameter space
as shown by the green contour.\  
For a smaller Hubble parameter $H_{\rm{inf}}\hsm\!=\hsm\! 10^{13}\hp$GeV,
our predictions of the local-type NG are beyond the reach of Planck-2018, 
but will be largely probed by the future measurements with improved sensitivities
of $f_{\rm{NL}}^{\rm{local}}\!=\!\pm 1,\,\pm 0.1,\,\pm 0.01$
(shown by the orange, yellow and white contours).\

We note that the low-energy neutrino oscillation data provide 
$\Delta m_{13}^{2}\!\simeq\! 2.5 \hsm\times\hsm 10^{-3}\,${eV}$^{2}$ and 
$\Delta m_{12}^{2}\!\simeq\! 7.4 \hsm\times\hsm 10^{-5}\,${eV}$^{2}$
\cite{nuMassFit}, 
requiring at least one of the light neutrinos has mass $m_\nu^{} \!\!\gtrsim\! 0.05\hp${eV}.\ 
Moreover, the cosmological measurements can place an upper bound on the sum of light neutrino masses.\ 
Combining this with the neutrino oscillation measurements on the mass-squared-differences 
can determine the upper limits of the light neutrino masses for either normal ordering (NO) 
or inverted ordering (IO).\ 
For instances, cosmological measurements based on the CMB alone already set a 95\% upper limit, 
$\sum\!m_\nu^{} \!\lesssim\! 0.26\hp${eV}\,\cite{Planck6}.\  
Combining this with the observations of large-scale structure, eBOSS Collaboration\,\cite{eBOSS:2020yzd} 
placed a 95\% upper bound $\sum\!m_\nu^{} \!\!\lesssim\! 0.10\hp${eV}
and DES Collaboration\,\cite{DES:2021wwk} set a constraint 
$\sum\!m_\nu^{} \!\!\lesssim\! 0.13\hp${eV} at 95\%\,C.L.\ 
%
%
Combining the tighter bound $\sum\!m_\nu^{} \!\!\lesssim\! 0.13\hp${eV} 
with the neutrino oscillation data\,\cite{nuMassFit},  
we find that the largest light neutrino mass to be $m_3^{}\!\hsm\simeq\!0.06\hp${eV} for the NO and 
$m_2^{}\!\hsm\simeq\!0.05\hp${eV} for the IO.\ 
Fig.\,\ref{fig:1} shows that the current and future measurements
of the local-type NG are sensitive to probing the difference between the cases with
light neutrino mass $m_\nu^{}\!\!=\! 0.06\hp$eV (pink curves) versus
$m_\nu^{}\!\!=\! 0.05\hp$eV (purple curves).\ 
The forthcoming oscillation experiments such as JUNO\,\cite{JUNO:2015zny} and DUNE\,\cite{DUNE:2016hlj} 
are expected to determine the neutrino mass ordering, and give stronger constraints on 
the allowed light neutrino masses.\ 
It is encouraging to see that using the NG measurements to probe neutrino seesaw
around the inflation scale (in Fig.\,\ref{fig:1}) 
could also have sensitivity to the light neutrino masses and their ordering.\ 
Hence this may interplay with the low-energy oscillation experiments such as JUNO and DUNE.\

\vspace*{1mm} 

Additionally, for the Higgs modulated reheating, 
the local NG also depends on the SM Higgs self-coupling $\lambda\,$, which can be used to probe the Higgs self-coupling.\
In order to test the sensitivity of $f_{\mathrm{NL}}^{\mathrm{local}}$ to the SM Higgs self-coupling $\lambda\,$, we vary the value of $\lambda$ and present the $f_{\mathrm{NL}}^{\mathrm{local}}$ contours at $2\sigma$ level in Fig.\,\ref{fig:2}.\
The $f_{\rm{NL}}^{\rm{local}}$ contours in solid (dashed) curves 
correspond to the Higgs self-coupling $\lambda \!=\! 0.01\,(0.02)$.\ 
The seesaw predictions are presented by pink curves for the light neutrino mass of 
$m_\nu^{}\!\!=\!0.05{\hp}$eV, with the Hubble parameter $H\!\!=\!\!10^{13}\hp$GeV and $H\!=\!3\times\!10^{13}\hp$GeV respectively.
In Fig.\,\ref{fig:2}, with a larger Higgs self-coupling value $\lambda\!=\!0.02$,
the non-Gaussianity contours (in dashed curves) impose weaker bounds 
on the seesaw parameter space of 
$(y_\nu^{},M)$ as compared to the contours (in solid curves) 
with a smaller coupling $\lambda\hsm\!=\!0.01\hp$.\footnote{%
This is in contrast with the conventional collider probe of the Higgs self-coupling $\lambda\,$,
where a larger $\lambda$ value always produces stronger signals of the di-Higgs production\,\cite{He:2015spf}.}\   
This analysis shows that the NG measurements of 
$f_{\mathrm{NL}}^{\mathrm{local}}$ are sensitive to the probe of the Higgs self-coupling $\lambda$
at the seesaw scale, which is quantitatively connected to the low-energy values of $\lambda$ 
(measured by the LHC and future high energy colliders\,\cite{He:2015spf}) via the renormalization group evolution.\ 
Hence, this also demonstrates the interplay on probing the Higgs self-coupling $\lambda$ 
between the high-scale cosmological NG measurements and the TeV-scale collider measurements.\ 

In Table\,\ref{tab:1}, we show the dependence of the NG 
on different values of the cutoff scale $\Lambda$ and 
inflaton mass $m_\phi^{}$.\ 
For illustration, we choose a sample input of neutrino seesaw scale $M\!\!=\!5H_{\rm{inf}}^{}$ (with $H_{\rm{inf}}^{}\hsm\!=\hsm\!10^{13}\hp$GeV), 
Higgs-neutrino Yukawa coupling 
$\,y_{\nu}^{}\hsm\!=\!0.5\hp$, and 
the SM Higgs self-coupling 
$\lambda \!=\! 0.01\hp$.\ 
Such sample inputs may indicate certain parameter degeneracy 
in NG signals.\ But the Hubble scale during inflation ($\Hinf$) could be determined from other measurements in principle (such as the tensor-to-scalar ratio).\ 
The mass of $N_{\!R}^{}$ 
may be inferred from the cosmological collider signatures, 
whereas the light neutrino masses can be measured 
by the on-going and future low-energy neutrino experiments.\ 
Although $\Lambda$ and $m_\phi^{}$ are more closely related 
to the UV physics and thus more challenging to determine, 
such degeneracy could be potentially resolved 
by analyzing the higher-point correlation functions
(such as the four-point function).\  
In principle, measuring more detailed information from 
higher-point correlators can help to further pin down 
the underlying parameter space and to distinguish our scenario from other models generating local non-Gaussianity.\

\begin{table}[t]
\centering

\begin{tabular}{c|cc|cc|cc|cc}
\hline\hline
Benchmarks &\, $A_1$\, & \,$A_2$\, & \,$B_1$\, & \,$B_2$\, &\, $C_1$ & \,$C_2$\, & $D_1$ & $D_2$ \\ 
\hline\hline
 $\Lambda/H_\text{inf}$ & $60$ &  $60$ &  $70$ &  $70$ &  $80$ &  $80$ &  $100$ &  $100$ \\ 
\hline
$m_\phi/H_\text{inf}$ & $30$ &  $40$ &  $30$ &  $40$ &  $30$ &  $40$ &  $30$ &  $40$ \\ 
\hline
$f_{\mathrm{NL}}^{\rm{local}}$& .34 & 1.8 & .10 & .54 & .034 & .18 & .01 & .031\\ 
\hline
\hline
\end{tabular}
\vspace*{-1mm}
\caption{Comparison of the non-Gaussianity predictions in our framework for three sets of benchmark points with specific  cutoff scales $\Lambda$ and inflaton masses $m_\phi$.\ 
}
\label{tab:1}
\vspace*{-5mm}
\end{table}

The ongoing and forthcoming measurements on the 
non-Gaussianity (such as those from  DESI\,\cite{DESI:2016fyo}, CMB-S4\,\cite{Abazajian:2019eic}, 
Euclid\,\cite{Euclid},  SPHEREx\,\cite{SPHEREx:2014bgr}, LSST\,\cite{LSST}, 
and SKA \cite{SKA} experiments) will further probe the origin of neutrino mass generation 
through the seesaw mechanism around inflation scale.\ 
We stress that  once the local non-Gaussinity is observed, 
the associate cosmological collider signals\,\cite{Arkani-Hamed:2015bza}\cite{Chen:2018xck}\cite{Hook:2019zxa} 
can be used to discriminate our scenario from other local non-Gaussinity sources.\  
A systematic expansion of this Letter is presented in the companion longer paper 
of Ref.\,\cite{Long}.

\vspace*{2mm}
\noindent 
\textbf{Acknowledgments}\\[1mm] 
We thank Xingang Chen, Misao Sasaki, Zhong-Zhi Xianyu, and Yi Wang for useful discussions.\  The research of H.\,J.\,H., L.\,S.\ and J.\,Y.\ was supported in part by the National Natural Science Foundation of China (Grant Nos.\,12175136, 12435005 and 11835005) and by the Shenzhen Science and Technology Program (Grant No.\ JCYJ20240813150911015).\ C.\,H.\ acknowledges support from the Sun Yat-Sen University Science Foundation, the Fundamental Research Funds for the Central Universities at Sun Yat-sen University under Grant No.\,24qnpy117, the National Key R{\&}D Program of China under Grant 2023YFA1606100, the National Natural Science Foundation of China under Grants No.\,12435005, and  the Key Laboratory of Particle Astrophysics and Cosmology (MOE) 
of Shanghai Jiao Tong University.\
This work was supported in part by the State Key Laboratory of Dark Matter Physics at Shanghai Jiao Tong University.


\begin{thebibliography}{99}
\bibitem{foot0}
Adding additional light scalar particle(s) to the Higgs sector at weak scale could lift 
the Higgs self-coupling to the level of ${O}(0.1)$ at the inflation scale\,\cite{HX}.\
For the current study we will just adopt the minimal SM Higgs sector.
\bibitem{foot-UVmodel}
As an illustration, this dimension-5 operator 
in Eq.\eqref{Total_Lagrangian} 
may be induced from a UV model having an approximate 
global $U(1)_{B-L}^{}$ symmetry 
which is spontaneously broken by a new scalar field $\Phi$ 
with a $U(1)_{B-L}^{}$ charge $-2$ and 
having a vacuum expectation value (VEV), 
$\left<\Phi\right>\hsm\!=\!\!f\hp$.\ 
After spontaneous symmetry breaking (SSB), 
the Yukawa interaction
$(y_x^{}\overline{N^c_R}N_R\Phi\!+\hsm\rm{H.c.})$ 
will generate a Majorana mass $M\!=\!y_x^{}f$
for the right-handed neutrino $N_R^{}$, where   
$f\!\!=\!\!M/y_x^{}\hsm\!\!=\!\!O(10)M\hp$ for a 
natural Yukawa coupling $y_x^{}\!=\!O(0.1)$.\  
The inflaton $\phi$ emerges as a pseudo-Nambu-Goldstone boson 
from $\Phi$ and has its mass generated by soft breaking
of the $U(1)_{B-L}^{}$ symmetry; and its interactions with fermions are dictated by their U(1) charges.\ 
The residual shift symmetry enforces that the inflaton-fermion coupling (such as the $\phi\,$-$N_R^{}$-$N_R^{}$ coupling) 
takes the form of a dimension-5 operator.\ 
Its cutoff scale $\Lambda\!=\!f\hp$ is given by the 
$U(1)_{B-L}^{}$ breaking scale $f$, which is about a factor 
of $O(10)$ of the $N_R^{}$ mass scale as shown above 
since the $N_R^{}$ mass originates from the SSB of 
$U(1)_{B-L}^{}$.
\bibitem{foot-x}
In Ref.\,\cite{Karam200614404}, the authors introduced a dimension-4 coupling between inflaton ($\phi$) 
and right-handed neutrino ($N_R^{}$), which generates the Majorana mass term for $N_R^{}$.\ 
They made an assumption that the Dirac mass   
$y_h^{}h\!\!\gg\!\! M$ with negligible Majorana mass $M$
of the right-handed neutrino.\ 
In this way, the inflaton decay is kinematically blocked
if the inflaton mass is smaller than twice of the Dirac mass.\ 
Once the Higgs vacuum expectation value decreases below a certain threshold, 
the inflaton can decay.\ 
This is a different scenario of Higgs modulated reheating and 
the resultant NG originates from this blocking effect.\ 
%
This scenario assumed negligible Majorana mass $M$ of $N_R^{}$ and did not test high scale seesaw.\ 
Besides, the dimension-4 coupling between inflton and $N_R^{}$ has to be generally suppressed 
to keep the flatness of inflaton potential (which is realized by imposing an approximate 
shift symmetry in our framework), so our dimension-5 coupling
in Eq.\eqref{Total_Lagrangian} plays the major role for inflaton decay.\ 
Hence, our present work has proposed a new scenario of 
Higgs modulated reheating and can test {high scale} seesaw mechanism,  
which would be complementary to the case of Ref.\,\cite{Karam200614404}.
\bibitem{foot2}
Here $t_\text{inf}^{}$ is the physical time at the end of inflation, 
$t_\text{reh}^{}$ is the physical time at which reheating occurs,
$\rho_{\rm{f}}^{}$ is a reference energy density and 
$t_{\rm f}^{}$ is the reference time where the energy density $\rho=\rho_{\rm{f}}^{}$
after the completion of reheating.\
\bibitem{foot3}
Here $\langle \delta h_{\mathbf{k_1}}\delta h_{\mathbf{k_2}}\delta h_{\mathbf{k_3}} \rangle'$ is defined as the 3-point correlation function without momentum delta function:\\  
$\langle \delta h_{\mathbf{k_1}}^{}\delta h_{\mathbf{k_2}}^{}\delta h_{\mathbf{k_3}}^{}\rangle
=(2\pi)^3\delta^3(\mathbf{k}_{1}+\mathbf{k}_{2}+\mathbf{k}_3)
\langle\delta h_{\mathbf{k_1}}^{}\delta h_{\mathbf{k_2}}^{}\delta h_{\mathbf{k_3}}^{}\rangle'$.
\bibitem{foot-nu}
These inputs correspond to the light neutrino mass $m_\nu^{}$ at the low-energy electroweak scale.\ 
The value of $m_\nu^{}$ at the seesaw scale or Hubble scale is connected to its low energy value
via renormalization-group (RG) running and is enlarged by about 30\% relative to
its low energy value\,\cite{Antusch0305273}\cite{He11042654}.\ 
The RG running effects are included in our present analysis.
\end{thebibliography}
\end{document}